\documentclass[twocolumn,twoside,slac_two]{revtex4}
\usepackage{graphicx}  
\usepackage{fancyhdr}
\begin{document}

\title{Simulation of Relativistic Jets and Associated Self-consistent Radiation}

%

\author{K.-I. Nishikawa}
\affiliation{National Space Science and Technology Center,
  Huntsville, AL 35805, USA}
\author{E. J. Choi, K. Min}
\affiliation{Korea Advanced Institute of Science and Technology, 
Daejeon 305-701, South Korea}

\author{P. Hardee}
\affiliation{Department of Physics and Astronomy,
  The University of Alabama, 
  Tuscaloosa, AL 35487, USA}

\author{Y. Mizuno}
\affiliation{Institute of Astronomy
National Tsing-Hua University,
Hsinchu, Taiwan 30013, R.O.C}

\author{B. Zhang}
\affiliation{Department of Physics, University of Nevada, Las
Vegas, NV 89154, USA}

\author{J. Niemiec}
\affiliation{Institute of Nuclear Physics PAN, ul. Radzikowskiego 152, 31-342 Krak\'{o}w, Poland}

\author{M. Medvedev}
\affiliation{Department of Physics and Astronomy, University of Kansas, KS
66045, USA}

\author{\AA. Nordlund, J. Frederiksen}
\affiliation{Niels Bohr Institute, University of Copenhagen, Juliane Maries Vej 30, 
2100 K\o benhavn \O,Denmark}

\author{M. Pohl}
\affiliation{Institue of Physics and Astronomy, University of Potsdam, 
14476 Potsdam-Golm
Germany}

\author{H. Sol}
\affiliation{LUTH, Observatore de Paris-Meudon, 5 place Jules Jansen, 92195 Meudon Cedex, France}

\author{D. H. Hartmann}
\affiliation{Department of Physics and Astronomy, Clemson University, Clemson, SC 29634, USA}

\author{G. J. Fishman}
\affiliation{NASA/MSFC,
  Huntsville, AL 35805, USA}

\begin{abstract}
Plasma instabilities excited in collisionless shocks are responsible for particle acceleration. We have investigated the particle acceleration and shock structure associated with an unmagnetized relativistic electron-positron jet propagating into an unmagnetized electron-positron plasma. Cold jet electrons are thermalized and slowed while the ambient electrons are swept up to create a partially developed hydrodynamic-like shock structure. In the leading shock, electron density increases by a factor of about 3.5 in the simulation frame. Strong electromagnetic fields are generated in the trailing shock and provide an emission site. These magnetic fields contribute to the electronÕs transverse deflection behind the shock. Our initial results of a jet-ambient  interaction with anti-parallel magnetic fields show pile-up of magnetic fields at the colliding shock, which may lead to reconnection and associated particle acceleration. We will investigate the radiation in transient stage as a possible generation mechanism of precursors of prompt emission. In our simulations we calculate the radiation from electrons in the shock region. The detailed properties of this radiation are important for understanding the complex time evolution and spectral structure in gamma-ray bursts, relativistic jets, and supernova remnants.

\end{abstract}

\maketitle

\thispagestyle{fancy}


\section{Introduction}
Particle-in-cell (PIC) simulations can shed light on the physical
mechanism of particle acceleration that occurs in the complicated
dynamics within relativistic shocks.  Recent PIC simulations of
relativistic electron-ion and electron-positron jets injected into an
ambient plasma show that acceleration occurs within the downstream 
jet~\cite{nishi03,fred04,nishi05,nishi06,ram07,anat08a,anat08b,sironi09m,nishi09a}.

In general, these  simulations have confirmed that relativistic jets excite 
the Weibel instability, which generates current filaments and associated
magnetic fields~\cite{weib59,medv99} and accelerates, 
electrons~\cite{hede05,sironi09m}.
Therefore, the investigation of radiation resulting from accelerated particles 
(mainly electrons and positrons) in turbulent magnetic fields is essential for 
understanding radiation mechanisms and their observable spectral properties. 

Recently, synthetic spectrum has been obtained using RPIC simulations~\cite{hedeT05,nishi09b,martins09,
sironi09j,fred10} in order to examine ``jitter radiation''~\cite{medv00,medv06}. Further investigations are required to understand radiation mechanisms for gamma-ray bursts and variabilities in radiation from AGN jets.

\begin{figure*}[t]
\centering
\includegraphics[width=125mm]{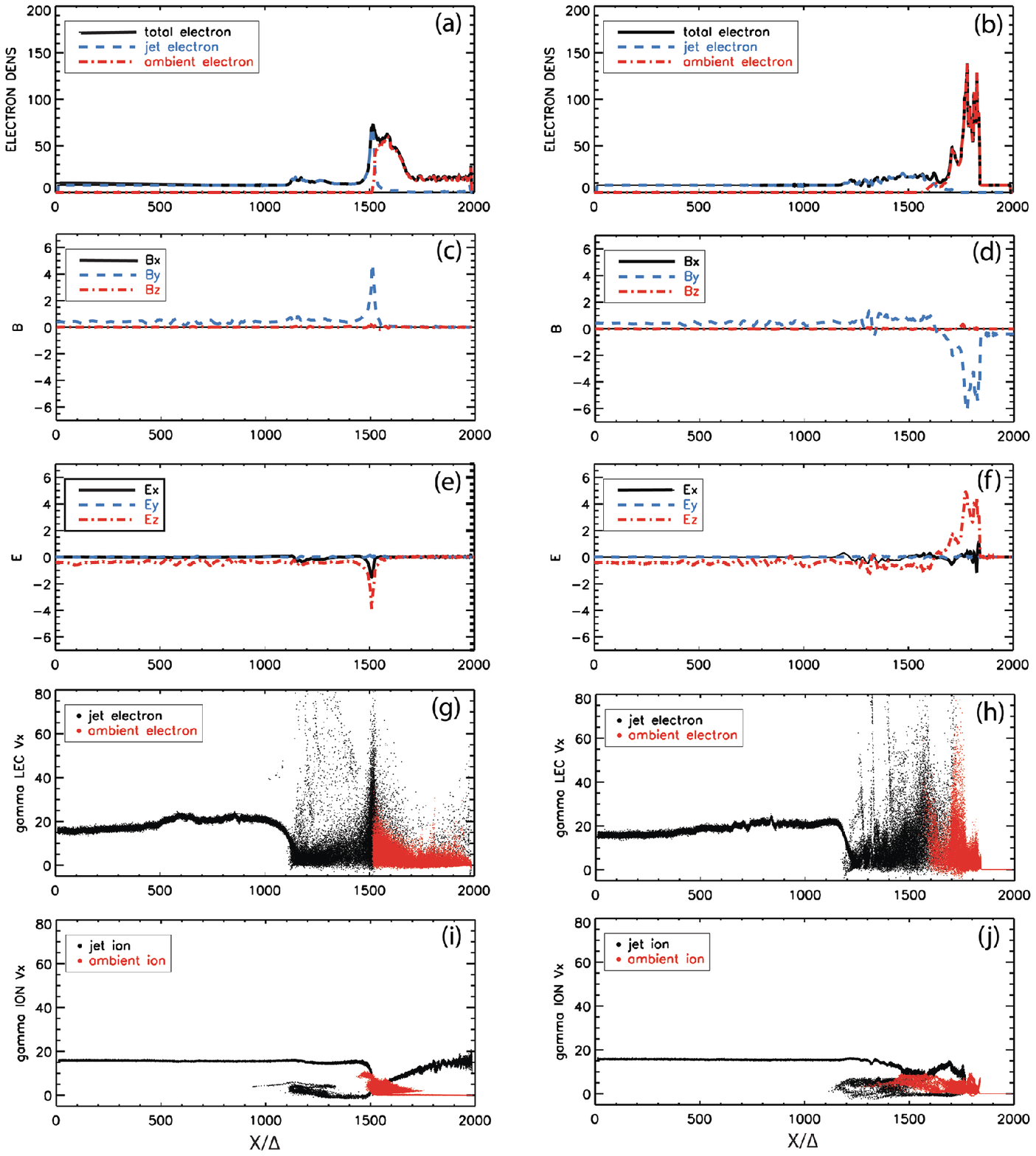}
\caption{Snapshots of the collision of a perpendicularly magnetized jet with an unmagnetized (left column), or anti-parallel magnetized (right column) ambient plasma  at simulation time $t = 1450\omega_{\rm pe}^{-1}$ (Choi, Min, \& Nishikawa 2011). The different panels show averaged values of the electron density (a) and (b), the magnetic field (c) and (d), the electric field (e) and (f), the phase space of electrons (g) and (h), and the phase space of ions (i) and (j). Reconnection occurs for the case involving anti-parallel magnetic fields and is indicated by the positive $E_{\rm y}$ component in Fig.\ 1f.} 
\end{figure*}
\section{Reconnections}

Recently, reconnection has been proposed for additional particle acceleration mechanism for AGN jets and gamma-ray burst jets~\cite{uzden11,granot11a,granot11,mck11,Zhang11,gian10,gian11b,komis09}.  Various reconnection simulations have been performed; RPIC simulations~\cite{dsught2010,zen11,fujim11,sironi11b}, resistive relativistic MHD (RRMHD)~\cite{komis07,zen10a,takah11}, and two-fluid~\cite{zen09a,zen09b}  simulations. In addition, the Kelvin-Helmholtz instability (KHI) may also lead to particle acceleration~\cite{Alves11}.

In order to investigate the evolution of ejecta and associated emission we  inject jets containing a perpendicular magnetic field and associated convective electric field ($E = -v_{\rm j} \times B_{\perp}$) varying
the magnetic field strength, i.e., magnetization parameter $\sigma = B_{\perp}^{2}/(4\pi n_{\rm j}\gamma m_{\rm e} c^{2})$~\cite{diec08,jin11}. These simulations are different from the previous simulations where jets were injected into a perpendicularly magnetized ambient plasma~\cite{hede05}. We have investigated the evolution of colliding magnetized shells and calculate radiation as has been done theoretically and for RMHD simulations in order to include self-consistent microscopic effects~\cite{mimcaM10}.

%
\begin{figure*}[t]
\centering
\includegraphics[width=105mm]{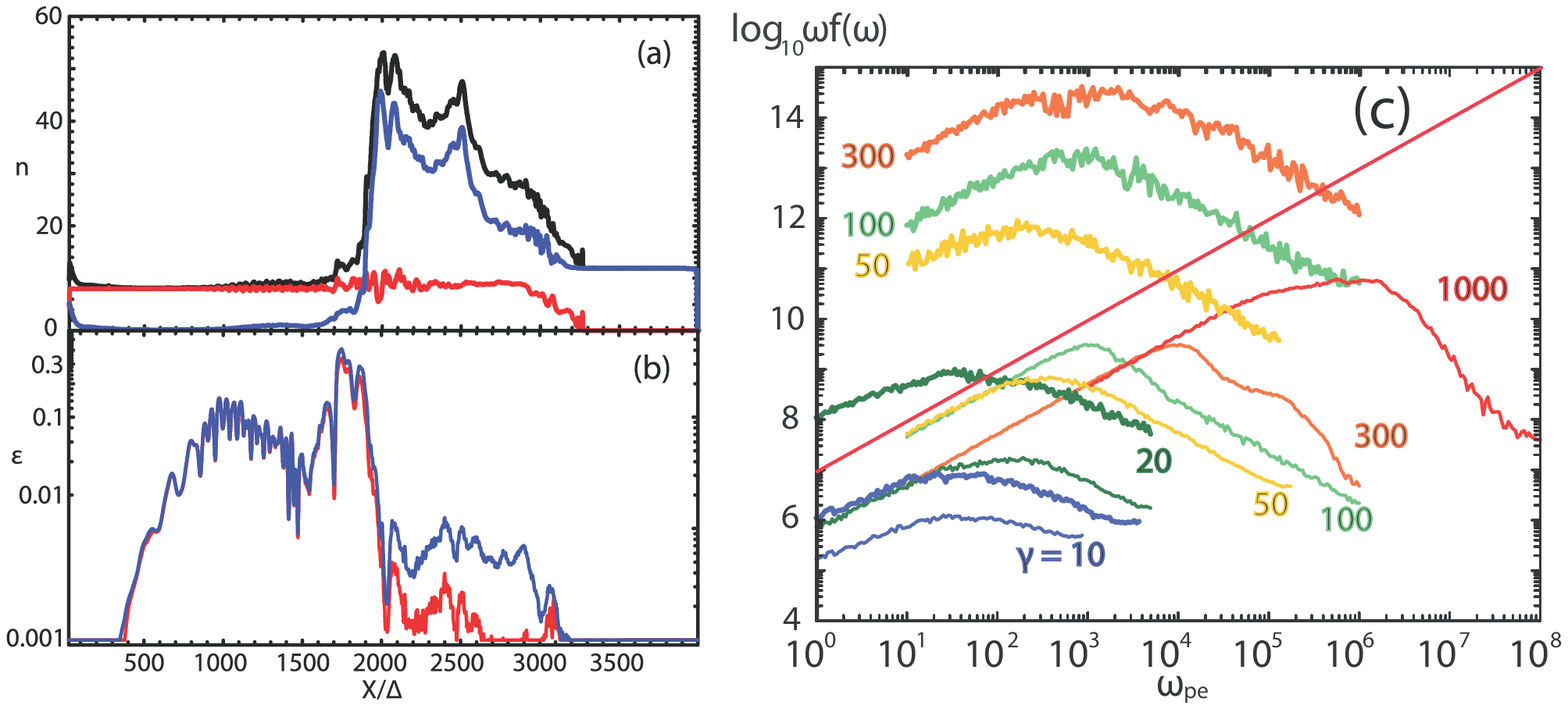}
\caption{Averaged values of (a): jet (red), ambient (blue), and total
(black) electron density, and (b): electric (red) and magnetic (blue)
field energy divided by the jet kinetic energy at $t = 3250~\omega_{\rm
pe}^{-1}$. (c): the spectra are for jets injected with Lorentz factors
of $\gamma =$ 10, 20, 50, 100, 300 and 1000 with cold (thin lines) and warm (thick lines) electrons. The low frequency slope is
approximately 1.
\label{fig1}}
\end{figure*}

\subsection{Recent RPIC Simulations with Magnetized Jets Colliding with Unmagnetized and Anti-Parallel
Magnetized Ambient Plasmas}

We have performed simulations using a system with ($L_{\rm x},
L_{\rm y}, L_{\rm z}) = (2005\Delta, 65\Delta, 65\Delta)$ and a 
total of  $\sim$ a few  million particles (8 particles$/$cell$/$species 
for the ambient plasma) in the active grid zones~\cite{jin11}. 
In the simulations the electron skin depth, $\lambda_{\rm ce} = c/\omega_{\rm pe} =
10.0\Delta$, where $\omega_{\rm pe} = (4\pi e^{2}n_{\rm e}/m_{\rm
e})^{1/2}$ is the electron plasma frequency and the electron Debye length 
$\lambda_{\rm e}$ is half of the grid
size. Here the computational domain is three times longer than in our
previous simulations~\cite{nishi09a}. 
The electron number density of the jet is as same as 
the ambient electron density and $\gamma  = 15$.  In this study the jets collide with
the ambient plasmas at $x = 500\Delta$.
The electron thermal velocity of jet is $v^{\rm e}_{\rm j,th} =
0.014c$, where $c = 1$ is the speed of light.  Radiating boundary conditions were used on the
planes at {\it $x = x_{\min}~{\&}~x_{\max}$}. Periodic boundary
conditions were used on all transverse boundaries. 
The ambient and jet electron-ion plasma has mass ratio $m_{\rm
i}/m_{\rm e} = 16$. 
The electron/positron thermal velocity in the ambient plasma is 
$v^{\rm e}_{\rm a,th} = 0.05c$ 
and the ion thermal velocity is $v^{\rm i}_{\rm th} = 0.022c$
where $c = 1$ is the speed of light.

As in previous papers~\cite{nishi09a}, 
the 
``flat" (thick) jet fills the computational domain in the transverse
directions (infinite width). Thus, we are simulating a small section
of a relativistic shock infinite in the transverse direction. 

Figure 1 shows snapshots of the shocks generated by a jet propagating into an ambient plasma at  simulation time $t = 1450\omega_{\rm pe}^{-1}$ with magnetization parameter $\omega^{2}_{\rm ce}/\omega^{2}_{\rm pe} =1$. Here the jet carries a $B_{\rm y}$ magnetic field component with convective electric field component $E_{\rm z}$. Panels in the left column show a case with no magnetic field in the ambient plasma, see 
the dotted blue line in Fig 1c. Panels in the right column show a case with an anti-parallel magnetic field ($-B_{\rm y}$) in the ambient plasma, see the dotted blue line in Fig. 1d.

The anti-parallel magnetic field in the ambient leads to dramatic evolution in the collision region
as shown on the right column. The electron density piles up at the jet front (Fig. 1b), negative strong $B_{\rm y}$ (Fig. 1d) and positive strong $E_{\rm z}$ (Fig. 1f) are found and indicate the occurrence of reconnection. In the case of no ambient
magnetic field, jet electrons and ions propagate through the collision region (Figs. 1g and 1i), as opposed to the anti-parallel magnetized ambient that hinders jet particle propagation through the ambient (Fig. 1b). In the relatively short simulation time, electrons are accelerated promptly and strongly. As shown in  Fig. 1, the magnetic fields play an essential role in particle acceleration and, of course, in the generation of 
radiation. In this proposal we will systematically investigate the effects of magnetic fields in relativistic flow collisions including reconnection. 

\section{Electron-Positron Jet and Synthetic Radiation}

Figures 2a \& b show the averaged (in the $y-z$ plane) jet (red), ambient (blue), and total (black) electron density and electromagnetic field energy divided by the total jet kinetic energy from one simulation~\cite{nishi09a}. 
The maximum density in the forward shocked region is about five times the initial ambient density.  The jet-particle density remains nearly constant up to near the jet front. Current filaments and strong electromagnetic fields accompany growth of the Weibel instability in the trailing shock region.  

The synthetic spectra shown in Figure 2c are obtained for emission from jets with Lorentz factors of $\gamma =$ 10, 20, 50, 100, 300 and 1000 with cold (thin lines) and warm (thick lines) electrons  (Nishikawa et al. 2011a,b,c,d,e).   The radiation from the jet electrons shows a Bremsstrahlung-like spectrum for the eleven cases.

However, it should be noted that  at higher frequency the spectral slopes in Figure 1c are less steep than for a Bremsstrahlung spectrum. This is due to the Lorentz factor spread of accelerated jet electrons and a resulting higher average Lorentz factor. Additional spectral extension to high frequency is due to electron scattering in the magnetic fields generated by the Weibel instability~\cite{nishi09b}. 

\section{Concluding Remarks}

The recent simulations of colliding jets into ambient plasma with ant-parallel magnetic fields show that
drastic evolution of the shock with piled-up magnetic field with possible reconnection. At the colliding shock
electrons are accelerated strongly, which may generate strong radiation. We will   investigate this interesting
evolution further including synthetic spectra.

\bigskip 
\begin{acknowledgments}
This work is supported by NSF-AST-0506719, AST-0506666, AST-0908040, AST-0908010, 
NASA-NNG05GK73G, NNX07AJ88G, NNX08AG83G, NNX 08AL39G, and NNX09AD16G. 
Simulations were performed at the Columbia and Pleiades
at the NASA Advanced Supercomputing (NAS) and Ember at the National Center for Supercomputing
Applications (NCSA) which is supported by the NSF. Part of this work was done while K.-I. N. was 
visiting the Niels Bohr Institute. Support from the Danish Natural Science Research Council is gratefully acknowledged. This report was started  during the program ÒParticle Acceleration in Astrophysical 
PlasmasÓ at the Kavli Institute for Theoretical
Physics which is supported by the National Science Foundation under Grant No. PHY05-51164.
\end{acknowledgments}

\bigskip 

\begin{thebibliography}{9}   



\bibitem[Alves et al.(2011)]{Alves11}
E. P. Alves, T.  Grismayer, S. F. Martins,  F. Fi\'{u}za, R. A. Fonseca, and L. O. Silva, 
 (arXiv:1107.6037) (2011).


\bibitem[Choi, Min, \& Nishkawa(2011)]{jin11}
 E.J. Choi,  K. Min, and K.-I. Nishkawa, 
{\it ApJ} in preparation (2011).

\bibitem[Daughton et al.(2011)]{dsught2010}
W. Daughton,  V. Roytershteyn, H. Karimabadi, L. Yin, B. J.  Albright, B.    Bergen, and K. J.    Bowers,
{\it  Physics Nature} 
DOI: 10.1038/NPHYS1965 (2011).

\bibitem[Dieckmann, Shukla, \& Drury(2008)]{diec08}
M. E.   Dieckmann, P.K. Shukla, and L.O.C. Drury, 
{\it ApJ} {\bf 675}  586  (2008).

\bibitem{fred04}
J.T. Frederiksen, C.B. Hededal, T.  Haugb\o lle,  and \AA. Nordlund,
 {\it ApJ}  {\bf 608}  L13 (2004).

\bibitem{fred10}
J. T. Frederiksen, T. Haugb\o lle, M. V. Medvedev, and \AA. Nordlund,
{\it ApJ} {\bf 722} L114 (2010).

\bibitem[Fujimoto(2011)]{fujim11}
K. Fujimoto,
{\it J. Compt. Phys.} {\bf 230} 8508 (2011).


\bibitem[Giannios(2010)]{gian10}
D. Giannios, 
{\it MNRAS} {\bf 408}  L46 (2010).

\bibitem[Giannios(2011)]{gian11b}
D. Giannios, 
{\it J. Phys: Conf. Ser}. {\bf 283} 012015 (2011).


\bibitem[Granot(2011)]{granot11}
J. Granot, 
{\it  MNRAS} submitted (arXiv:1109.5314) (2011).

\bibitem[Granot et al.(2011a)]{granot11a}
J. Granot,  S. S. Komissarov, and A. Spitkovsky, 
{\it MNRAS} {\bf 411} 1323 (2011).

\bibitem{hedeT05}
C.B. Hededal,     Ph.D. thesis, (2005) (arXiv:astro-ph/0506559).


\bibitem[Hededal \& Nishikawa(2005)]{hede05} 
C. B. Hededal,  and  K.-I. Nishikawa, 
{\ it ApJ} {\bf 623} L89  (2005).



\bibitem[Komissarov(2007)]{komis07}
S. S. Komissarov, 
{\it MNRAS} {\bf 382} 995 (2007).

\bibitem[Komissarov et al.(2009)]{komis09}
S. S. Komissarov,  N. Vlahakis, A.
K\"onigl, and M. V.   Barkov, 
 {\it MNRAS} {\bf 397} 1153 (2009).

\bibitem{martins09}
J.L.  Martins, S.F. Martins,  R.A. Fonseca,  and L.O. and Silva, 
{\it Proc. of SPIE} {\bf 7359}  73590V-1 (2009).

\bibitem[McKinney \& Uzdensky(2011)]{mck11}
J. C. McKinney, D. A. Uzdensky, 
{\it MNRAS} online: 2 NOV 2011
DOI: 10.1111/j.1365-2966.2011.19721.x  (arXive:1011.1904) (2011).

\bibitem{medv99}
M.V. Medvedev,  and A. Loeb,  {\it ApJ} {\bf 526} 697 (1999).

\bibitem{medv00}
 M. V. Medvedev, 
{\it ApJ} {\bf 540}  704 (2000).

\bibitem{medv06}
M.V. Medvedev,   {\it   ApJ} {\bf 637}   869 (2006).

\bibitem[Mimca et al.(2010)]{mimcaM10}
P.  Mimica, D.  Giannios, and M. A. Aloy, 
{\it MNRAS} {\bf 407} 2501 (2010).

\bibitem{nishi03}  K.-I. Nishikawa, P. Hardee,
G. Richardson, R. Preece, H., Sol, and G.J. Fishman, {\it ApJ}
{\bf 595}  555 (2003).

\bibitem{nishi05}
K.-I. Nishikawa, P.   Hardee, G. Richardson, R. Preece, R., H. Sol, 
and G.J.  Fishman, {\it  ApJ} {\bf  623}  927 (2005).

\bibitem{nishi06} K.-I. Nishikawa,  P. Hardee,
C.B.  Hededal,  and G.J.  Fishman,  {\it ApJ} {\bf 642} 1267 (2006).

\bibitem{nishi09a}
K. -I. Nishikawa, J. Niemiec, P. Hardee,  M. Medvedev, H. Sol, Y. Mizuno, 
B. Zhang, M.  Pohl, M., Oka, and D.H.  Hartmann,  {\it  ApJ} {\bf 689} L10 (2009).

\bibitem{nishi09b}
K.-I.  Nishikawa, J. Niemiec, H.  Sol, M.  Medvedev, B. Zhang, \AA. Nordlund, J.T.
Frederiksen, P. Hardee, Y, Mizuno, D.H. Hartmann, and G.J.
Fishman,  
AIPC,  {\bf 1085} 589 (2009).

\bibitem{ram07}
 E. Ramirez-Ruiz, K.-I. Nishikawa,  and C.B. Hededal, {\it ApJ}
{\bf 671} 1877  (2007). 


\bibitem{sironi09m}
L.  Sironi,  and A.  Spitkovsky,   {\it ApJ} {\bf  698} 1523 (2009).

\bibitem{sironi09j}
L. Sironi,  and A.  Spitkovsky,  {\it ApJ}  {\bf 707} L92 (2010).

 \bibitem[Sironi \& Spitkovsky(2011b)]{sironi11b}
L.  Sironi, and A.  Spitkovsky, 
{\it  ApJ} {\bf 741} 39 ( 2011).

\bibitem{anat08a}
A. Spitkovsky,   {\it ApJ} {\bf 673} L39 (2008).

\bibitem{anat08b}
A. Spitkovsky,  {\it ApJ} {\bf 682} L5 (2008).

\bibitem[Takahashi et al.(2011)]{takah11}
H. Takahashi, T.  Kudoh, Y. Masada, and J. Matsumoto, 
{\it  ApJ} {\bf 739}:L53 (5pp) (2011).


\bibitem[Uzdensky(2011)]{uzden11}
 D. A. Uzdensky, 
{\it Space Sci. Rev.}
DOI 10.1007/s11214-011-9744-5 (2011).

\bibitem{weib59}
E.S. Weibel,  {\it Phys. Rev. Lett.} {\bf 2} 83 (1959). 

\bibitem[Zenitani et al.(2009a)]{zen09a}  
S. Zenitani, M. Hesse, and A.  Klimas, 
{\it  ApJ} {\bf 696} 1385 (2009a).
	 
 
\bibitem[Zenitani et al.(2009b)]{zen09b} 	
S. Zenitani, M. Hesse, and A.  Klimas, 		
{\it  ApJ} {\bf 705} 907 (2009b). 


\bibitem[Zenitani et al.(2010a)]{zen10a} 
S. Zenitani, M. Hesse, and A.  Klimas, 
{\it  ApJ} {\bf 716} 214 (2010).

\bibitem[Zenitani et al.(2011)]{zen11}
S. Zenitani, M. Hesse,  A.  Klimas, and M.   Kuznetsova,
{\it PRL} {\bf 106} 195003 (2011).

\bibitem[Zhang \& Yan(2011)]{Zhang11}
B. Zhang, and H. Yan, 
{\it ApJ} {\bf726}:90(23pp) (2011).

\end{thebibliography}

\end{document}